\documentclass[amsmath,amssymb,lineno, superscriptaddress, reprint,prl]{revtex4-1}
\usepackage{graphicx} 
\usepackage{dcolumn} 
\usepackage{bm} 
\usepackage{hyperref} 
\usepackage{longtable} 
\usepackage[T1]{fontenc}
\usepackage{times}
\usepackage{lipsum}
\usepackage{xcolor}
\usepackage{enumitem}
\graphicspath{{figures/} }

\begin{document}

\title{{Pressure--Tunable Photonic Band Gaps in an Entropic Colloidal Crystal} }

\author{Rose K.\ \surname{Cersonsky} }
\affiliation{Macromolecular Science and Engineering Program, University of Michigan, Ann Arbor MI 48109, USA}
\author{Julia \surname{Dshemuchadse} }
\affiliation{Department of Chemical Engineering, University of Michigan, Ann Arbor MI 48109, USA}
\author{James \surname{Antonaglia} }
\affiliation{Department of Physics, University of Michigan, Ann Arbor MI 48104, USA}
\author{Greg \surname{van Anders} }
\affiliation{Department of Physics, University of Michigan, Ann Arbor MI 48104, USA}
\author{Sharon C.\ \surname{Glotzer} }
\affiliation{Macromolecular Science and Engineering Program, University of Michigan, Ann Arbor MI 48109, USA}
\affiliation{Department of Chemical Engineering, University of Michigan, Ann Arbor MI 48109, USA}
\affiliation{Department of Physics, University of Michigan, Ann Arbor MI 48104, USA}
\affiliation{Department of Materials Science and Engineering, University of Michigan, Ann Arbor MI 48109, USA}
\affiliation{Biointerfaces Institute, University of Michigan, Ann Arbor MI 48109, USA}

\date{\today}

\begin{abstract}
Materials adopting the diamond structure possess useful properties in atomic and colloidal systems, and are a popular target for synthesis in colloids where a photonic band gap is possible. The desirable photonic properties of the diamond structure pose an interesting opportunity for reconfigurable matter: can we create a colloidal crystal able to switch reversibly to and from the diamond structure?
Drawing inspiration from high-pressure transitions of diamond-forming atomic systems, we design a system of polyhedrally-shaped particles that transitions from diamond to a tetragonal diamond derivative upon a small pressure change. The transition can alternatively be triggered by changing the shape of the particle \textit{in-situ}. We propose that the transition provides a reversible reconfiguration process for a potential new colloidal material, and draw parallels between this transition and phase behavior of the atomic transitions from which we take inspiration.
\end{abstract}

\maketitle

In the century since it was first characterized in 1913 by W.\ H.\ and W.\ L.\ Bragg, the diamond structure has been a popular focus of materials research \cite{Bragg1913}.
Diamond-type materials, including the diatomic equivalent, zincblende, exhibit extraordinary properties, such as high hardness, high thermal conductivity, and a high refractive index at room temperature. 
Also known for its optical properties, diamond was the first structure calculated to have a photonic band gap \cite{Ho1990}, meaning it will reflect all electromagnetic waves for a defined range of wavelengths.

The unique properties of diamond-structured materials provide an interesting target for reconfigurable materials:  a material designed to switch to and from a diamond-like structure with its concomitant properties.
There is precedent for reconfigurable diamond-structured materials in elemental structures: Si, Ge, and Sn each form the diamond structure at ambient pressure and transition to a tetragonal metallic phase, $\beta$-Sn, at higher pressures. However, these transitions are not reversible; both Si and Ge revert to metastable phases upon depressurization and cannot recover diamond \cite{Ackland2001,Clark1994}. 

Despite significant study into the underlying physics of these pressure-induced transitions \cite{Mcmillan2005,Jamieson1963,Bundy1996,Katzke2006,Ackland2001a}, few direct observations have been published on the transition mechanism, often only reporting conditions where this transition is absent, such as in zero temperature calculations \cite{Ackland2001a,Bundy1996,Jamieson1963, Goncharov1987}.
Why do diamond-type materials transition to the $\beta$-Sn structure under isotropic compression? Can an analogous structural transition be observed in diamond-forming colloidal crystals \cite{Zhang2005,Ducrot2017,Damasceno2012d,Jain, Wang, Romano2010}, where particles on the scale of nanometers to microns play the role of atoms? And if so, is a reversible transformation possible for colloidal crystals, or do metastable structures intercede?

Here we report a reversible phase transition for a simulated system of hard, convex particles with tetrahedral symmetry ordering via entropy maximization into a diamond structure. The transition is driven by pressure, and leads from the well-known cubic diamond structure to a tetragonal diamond derivative (TDD) distinct from the aforementioned $\beta$-Sn.The transition provides a reversible reconfiguration process for a potential new photonic colloidal material that can be switched among as many as four possible states, each of which permits a different frequency range of light in the infrared or visible regimes. Additionally, by noting that all three phases -- cubic diamond, TDD, and $\beta$-Sn -- are related through scaling along a major axis, we gain insight into the mechanism driving high-pressure transitions between diamond and $\beta$-Sn in atomic systems. 


\begin{figure}
\includegraphics[width=0.45\textwidth]{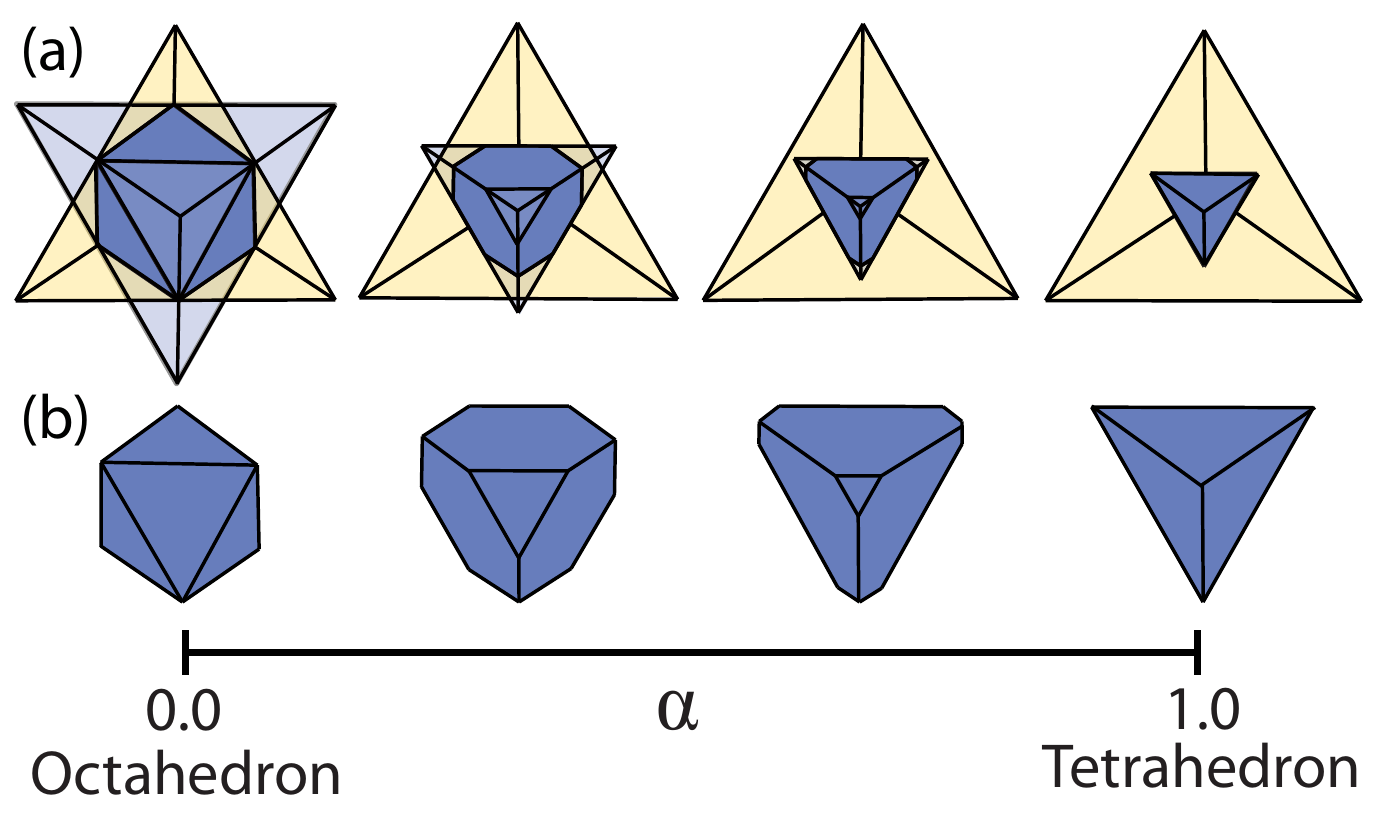}
\caption{\textbf{The $\Delta_{323}$ Shape Parameterization.} \textbf{(a--b)} The $\Delta_{323} $ shape family, ranging from an octahedron ($\alpha = 0.0$) to a tetrahedron ($\alpha = 1.0$).
 The intersection of planes at parameterized distances shown in \textbf{(a)} generate the resulting shapes in \textbf{(b)}.
 All shapes are scaled to unit volume in simulations.}
\label{fig:family323}
\end{figure}

We simulated a family of convex shapes with tetrahedral symmetry (here denoted $\Delta_{323} $) \cite{Chen2014} using the Digital Alchemy (DA) framework \cite{vanAnders2015}.
In DA, a design parameter is introduced as a thermodynamic state variable, which defines an extended \textit{alchemical ensemble} in which either the design parameter, the \textit{alchemical variable}, or its conjugate, the \textit{alchemical potential} can fluctuate while the other remains fixed.

The $\Delta_{323}$ family is constructed through the intersection of planes arranged tetrahedrally around the particle center, parametrized by $\alpha_3$, which can be viewed as the truncation along a set of three-fold axes, in the following called $\alpha$ for brevity.
The shape family is shown in Fig.~\ref{fig:family323}; all shapes are scaled to unit volume.
Shapes with $\alpha = 0.3\text{--} 0.5$ have been shown to assemble into diamond at densities $\phi = 0.5\text{--} 0.6$ \cite{Damasceno2012d}. At higher densities, we observe a transition to a tetragonal diamond derivative (TDD) structure.

\begin{figure} [htbp!]
\vspace{-10pt}
\includegraphics[width=0.45\textwidth]{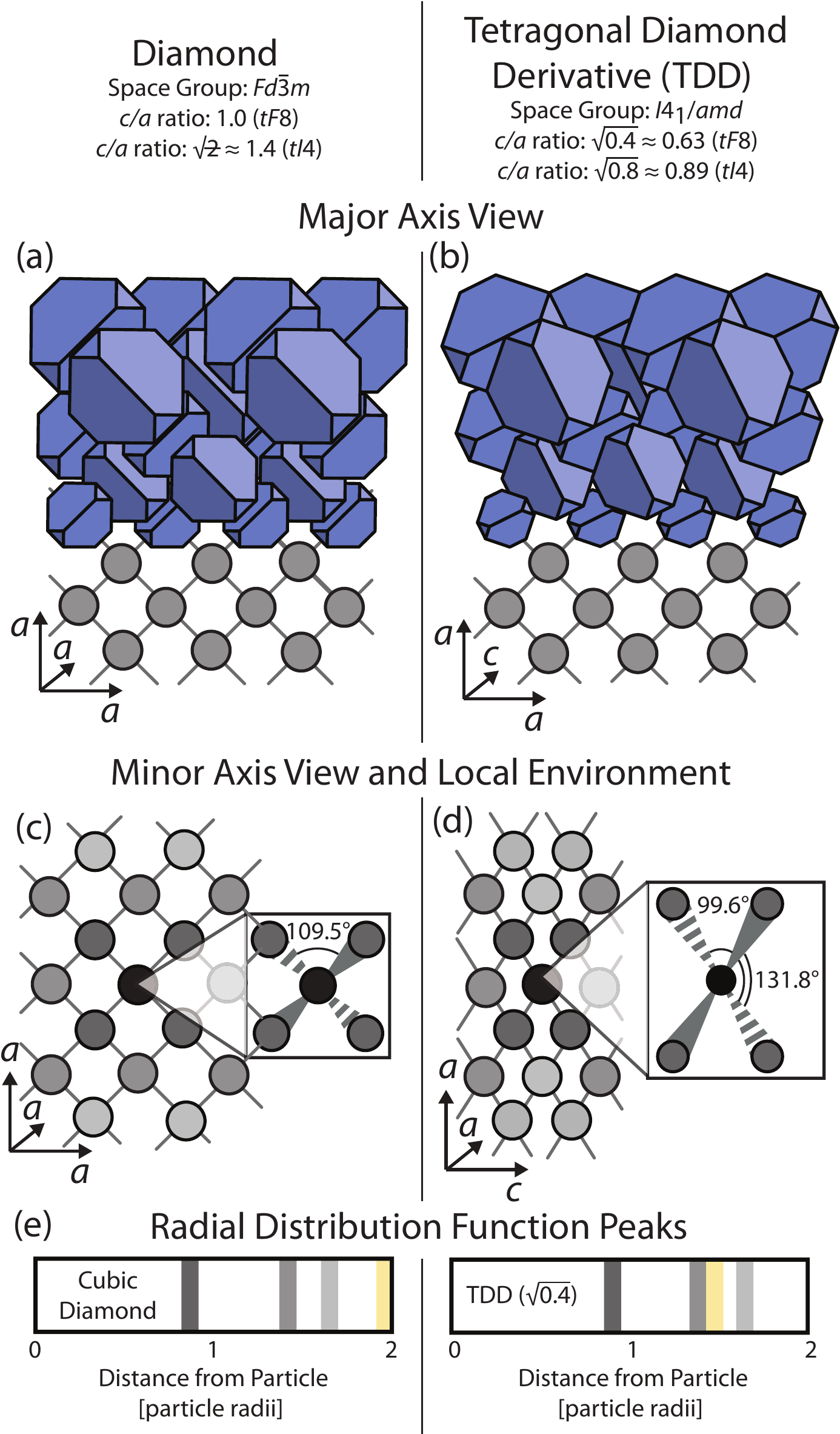}
\caption{\textbf{Structural Diagrams for Diamond and its Tetragonal Derivative.} \textbf{(a--d)} Structural comparison between \textbf{(a, c)} diamond and \textbf{(b, d)} the tetragonal diamond derivative (TDD) with $c/a = \sqrt{0.4}$, both shown with shape at $\alpha = 0.5$.
 Here, we show projections along the major axis of cubic diamond and along two differing axes in the tetragonal derivative structure ($\vec{c}$ in \textbf{(b)} and one of the equivalent $\vec{a} $-axes in \textbf{(d)}).
 The particle positions appear unchanged when viewed along the $\vec{c} $-axis, but the particles rotate about the $\vec{c} $-direction.
 \textbf{(e)} The peaks in the radial distribution function (RDF), \textit{i.e.}, the distances of the nearest-neighbor shells. The fourth nearest neighbors (yellow) in the $\vec{c} $-direction in diamond are immediately outside the third neighbor shell in TDD at $c/a = \sqrt{0.4} $.\vspace{-20pt}}
\label{fig:structs}
\end{figure}

These two crystal structures, diamond and TDD, differ in two major respects: the aspect ratio of the unit cell and the particle orientations.
The TDD structure is defined by lattice vectors $\vec{a} _1 = [a,0,0]$, $\vec{a} _2 = [0,a,0]$, $\vec{c} = [0,0,c]$, with particles situated at positions 0,0,0 and $\frac{1} {4},\frac{1} {4},\frac{1} {4} $ (in fractional coordinates) on a face-centered lattice
 (\textit{i.e.}, with centering vectors $[0,0,0],\,[\frac{1} {2},\frac{1} {2},0],\,[\frac{1} {2},0,\frac{1} {2} ]$, and $[0,\frac{1} {2},\frac{1} {2} ]$).
 For ease of comparison, tetragonal unit cells will be expressed in a non-standard face-centered setting $tF8$, such that all unit cells can be described in terms of $c/a$ for a face-centered unit cell (containing 8 particles), where $c/a = 1$ for the face-centered cubic unit cell of diamond ($cF8$-C).
Shapes from the $\Delta_{323} $ family densely pack into TDD with $c/a = \sqrt{0.4} \approx 0.6325$ (which is also $c/a_\textrm{min}$ for $\Delta_{323}$) and $\alpha = 0.5$.

The $\beta$-Sn structure (referred to by its Pearson symbol $tI4$-Sn) can be defined as a TDD structure at $c/a \approx 0.3859$ (given a $tF8$ unit cell).
$tI4$-Sn and TDD structures in $\Delta_{323}$ both have space group $I4_1/amd$ and Wyckoff position $4c$ $0,0,0$, albeit at different ratios of $c/a$.
TDD structures with $c/a$ $\geq \sqrt{0.4} $ have tetrahedral coordination (coordination number 4), whereas $tI4$-Sn has two additional particles adjacent to the first neighbor shell, resulting in an octahedral coordination (coordination number 6).
Other TDD phases have been found in studies of atomic structures although with aspect ratios near that of $tI4$-Sn or high-pressure $tI4$-Cs, which has $c/a > 1$ \cite{Bundy1996,Jamieson1963}. 

Snapshots of the two structures observed in our simulations are shown in Fig.~\ref{fig:structs}\textbf{(a--d)}, with their radial distribution function (RDF) peaks in Fig.~\ref{fig:structs}\textbf{(e)}.
Similar structure diagrams comparing diamond and TDD ($c/a=\sqrt 0.4$) with $tI4$-Sn can be found in the Supplementary Information.


We report phase diagrams comparing the $c/a$ ratio of the $tF8$ unit cell for varying densities and shape parameter $\alpha$. These phase diagrams reveal that, although all shapes with $0.3 \leq \alpha \leq 0.5$ assemble into diamond, they will transition to tetragonal unit cells at higher densities. The $NVT\alpha$ simulations generating this phase diagram and $NPT\alpha$ simulations for computing the equation of state were conducted by initializing both in the assembled and densely packed structures and equilibrating at the desired density or pressure.

\begin{figure} [htbp]
\includegraphics[width=0.48\textwidth]{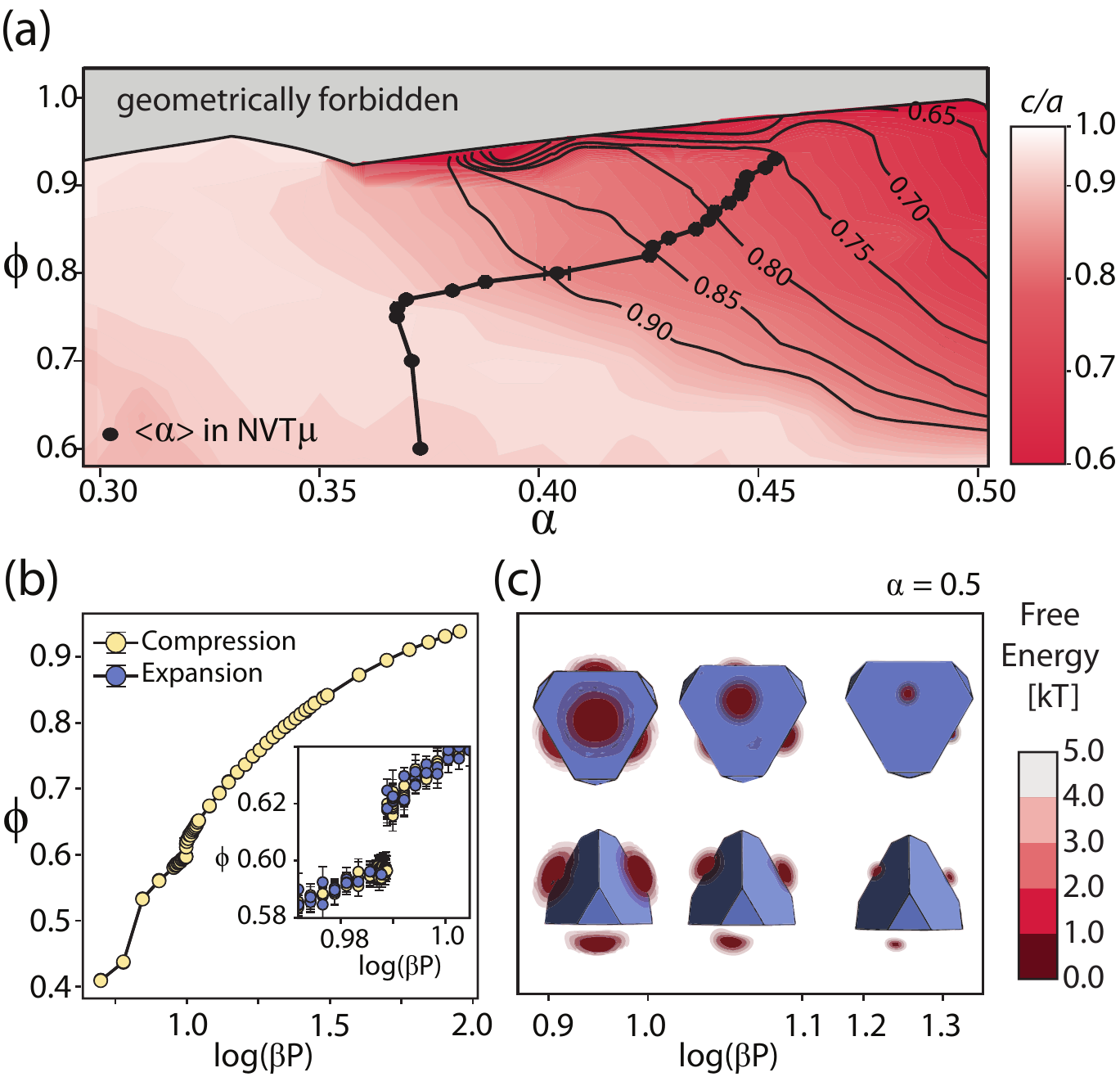}
\caption{\textbf{Simulation results for $NVT\alpha$, $NPT\alpha$, and $NVT\mu$.} \textbf{(a)} Composite ``phase diagram'' from all simulations.
 The putative densest packing in either diamond or TDD structure as a function of shape is the upper limit of the phase diagram, the geometrically forbidden region shown in grey. Shades of red represent the $c/a$ ratio of the unit cell, with the lighter red region showing where diamond is the configuration with the lowest free energy ($c/a=1$), and darker reds indicating TDD structures with lower $c/a$ values. The black data set represents results from $NVT\mu$ simulations, denoting  where the crossover from the diamond to the TDD stability region occurs as a function of density and shape.
 \textbf{(b)} Equation of state for $\alpha = 0.5$ evaluated in compression and expansion runs. The main plot shows the average values, with an inset showing the average and standard deviations around the transition pressure.
 \textbf{(c)} Potentials of mean force and torque for $\alpha = 0.5$ at varying pressures. The PMFT wells shift from the center of the large face of each particle to one edge. This shows that the coordination remains tetrahedral, but becomes distorted with increasing pressure.}
\label{fig:panel2}
\end{figure}

In Fig.~\ref{fig:panel2}\textbf{(a)}, the contour lines map the $c/a$ value of the unit cell of the equilibrium structure for systems of constant shape at varying densities. The shades of red represent the $c/a$ ratio of the unit cell lattice vectors, with the lighter red region showing where cubic diamond is the configuration with the lowest free energy ($c/a=1$), and darker reds indicating TDD structures, where $c/a<1$ is lower in free energy.
The region marked as geometrically forbidden (grey) shows the densities and $\alpha$ values at which particles would be forced to overlap.

In this study, shape and structure are intrinsically intertwined due to the shape-dependency of the transition from cubic diamond to its tetragonal derivative.
Using Alchemical hard particle Monte Carlo (Alch-HPMC) (described in \cite{Cersonsky2017} and in the Supplementary Information) with constant alchemical potential $\mu=0$, we can find the shape that produces the lowest free energy structure at a given density or pressure, and determine where the crossover from the diamond to the TDD stability region occurs ($\langle\alpha\rangle$ shown in black on Fig.~\ref{fig:panel2}\textbf{(a)}).
At low densities, the lowest free energy phase is diamond, but at densities $\gtrsim 0.75$ (or equivalently at unitless pressures $p^*\geq 10$--$15$) a TDD structure will have lower free energy in this shape space. Decompression of this high-density system showed full recovery of the diamond structure.

We investigated the nature of the phase transition and behavior of potentials of mean force and torque (PMFTs) for the phase transition corresponding to the largest change in $c/a$, at $\alpha = 0.50$, in Fig.~\ref{fig:panel2}\textbf{(b, c)}. In Fig.~\ref{fig:panel2}\textbf{(b)}, the main plot represents the average for both initializations, with an inset representing the average and standard deviations around the transition pressure. The transition exhibits a small discontinuity in $\beta P(\phi)$ and no hysteresis, which implies that it is either a weak first-order or a second-order transition \cite{Ehrenfest}. PMFTs, the calculation of which is detailed in \cite{VanAnders2014d}, provide a statistical landscape for entropic particle bonding---\textit{i.e.}\! they show the probable locations for the neighboring particles of a given reference particle in units of free energy. PMFTs show that after the transition occurs, the bonding remains four-fold, with the coordination being that of a distorted tetrahedron.


\begin{figure*}
\includegraphics[width=0.95\textwidth]{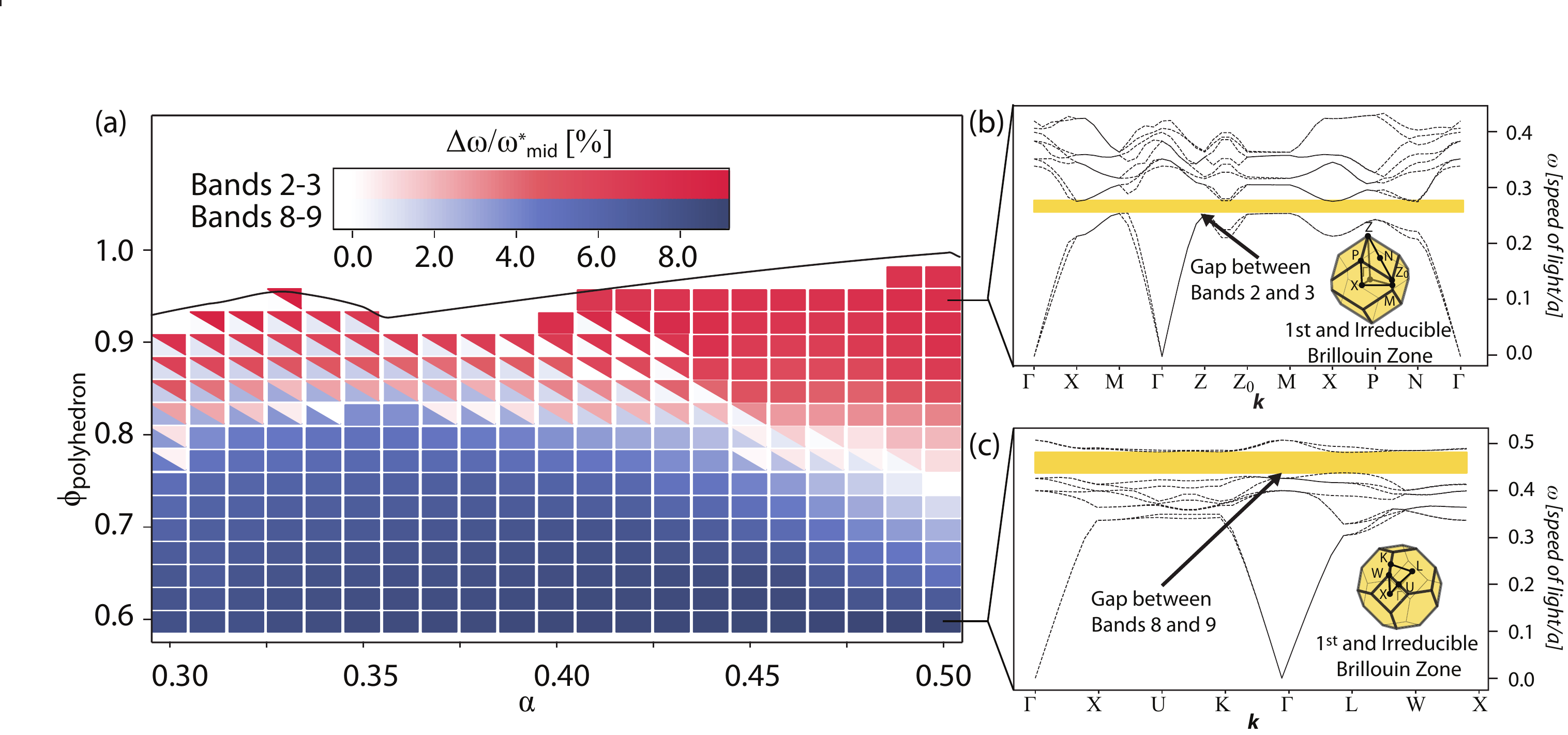}
\caption{\textbf{Photonic Band Gap Structures and Sizes} \textbf{(a)} Size of photonic band gaps between bands 2 and 3 (red) and bands 8 and 9 (blue) for structures in Fig.~\ref{fig:panel2}\textbf{(a)}. All polyhedra have been replaced with spheres with radius equal to the polyhedral insphere radius, and with dielectric constant $\epsilon = 11.56$. \textbf{(b--c)} Representative photonic band structures for $\alpha = 0.5$ at \textbf{(b)} $\phi=0.6$, where it is in cubic diamond, and (c) $\phi=0.95$, where it is in TDD at $\approx\sqrt{0.4}$. The insphere radius of $\Delta_{323} | _{\alpha=0.5}$ is 0.425.}
\label{fig:photonic}
\end{figure*}

Diamond exhibits a complete photonic band gap when the lattice sites are populated with dielectric spheres \cite{Ho1990}; thus the transition from diamond to any form of TDD would result in a shift in the photonic properties due to the symmetry reduction.
We computed the photonic band structure using MIT Photonic Bands (MPB)\cite{Johnson2001}, replacing each polyhedron with its insphere, since the calculation of photonic band structures for polyhedral particles is not yet available.
Although calculated using spherical geometry, these results are still pertinent for colloidal systems, as there exist synthesis techniques capable of creating polyhedral particles with spherical cores, either as core-shell colloids \cite{Li2017} or with polyhedral particle cages with low dielectric constant encasing a high dielectric constant sphere (e.g. \cite{Liu2016}).

We report the photonic band frequencies in units of $\text{(speed of light)}/a$, where $a$ is the lattice constant. Each complete photonic band gap is reported in unitless dimensions, defined as the width of the band gap over the mid-gap frequency, $\omega^*$. The conversion from $\omega^*$ to absolute wavelength is:\vspace{-10pt}

\begin{equation}
\lambda = \frac{a}{\omega^*}.
\label{eq:PBG}
\end{equation}

\vspace{-5pt}While we anticipated a change in photonic properties from an ``on'' to ``off'' state upon converting the diamond structure to TDD, we were surprised to find that the results are more consistent with a multistate material with one of four possibilities with regards to a complete photonic band gap: i) no gap, ii) a complete gap between bands 8 and  9, iii) a complete gap between bands 2 and 3, and (iv) complete gaps between both bands 2 and 3 and 8 and 9. Examples of (ii) and (iii) are shown in Fig.~{\ref{fig:photonic}\textbf{(b)} and \textbf{(c)}, respectively. These findings suggest how to make a material that could switch between permitting all light and blocking one set of frequencies or another.

Replacing each polyhedron with an insphere of dielectric constant $\epsilon = 11.56$ (that of silicon), we calculated the photonic band structure for the structures found in simulation for $\phi = 0.6$--$0.975$ (where $\phi$ is the density of the polyhedra) and $\alpha = 0.3$--$0.5$. The maximum band gap width was 9.2\% and all bands were centered around mid-gap frequencies of approximately 0.260--0.298 (between bands 2 and 3) or 0.384--0.459 (between bands 8 and 9). Unitless band gap widths are shown for $\Delta_{323}$ as a function of density in Fig.~{\ref{fig:photonic}\textbf{(a)}.

In addition to computing the photonic band structure for the dielectric constant of silicon, we also computed the photonic band gap widths for lower dielectric constants, included in the Supplementary Information. From these, we learn that complete photonic band gaps between bands 2 and 3 are possible with a minimum dielectric constant of 5.0, and between 8 and 9 with a minimum dielectric constant of 8.0. These values suggest that this type of photonic band structure would not be possible with crystals made of polymeric materials (which generally have $\epsilon < 4.0$), but would require materials of higher dielectric constant, such as silicon or germanium.

The phase transition we report at the colloidal length scale may also reveal insights about the atomic transition on which it is based. Although other TDD structures have not been observed as an intermediary between atomic diamond and $\beta$-Sn phases, such a consideration could provide new perspective into the mechanism underlying the phase transition.
In our simulations, all effects other than shape entropy are excluded \cite{VanAnders2014d}, and we restrict the particle geometry to symmetric truncations of tetrahedra, \textit{i.e.}, the $\Delta_{323} $ shape family.
Thus we prove that entropy alone suffices to explain the transition in colloidal systems from diamond to a lower $c/a$ tetragonal derivative. This correlation strengthens a supposition made by G.\ J.\ Ackland, who proposed that it was plausible that entropy is the driving force for diamond-to-$\beta$-Sn transitions, based on the absence of such a transition at 0 K \cite{Ackland2001}.

Our colloidal system differs from atomic systems in several respects. At the higher $c/a$ ratio ($\approx 0.63$ \emph{vs.}~$\approx0.39$), the coordination remains four-fold, while the neighboring particles form a distorted tetrahedron (as compared to a regular tetrahedron in perfect diamond). This retention of four-fold coordination may explain the reversibility in our system, compared to the irreversibility in atomic systems, which transition to octahedral coordination.

The same cannot be said for systems adopting the diatomic analog to diamond, zincblende, as these systems do not adopt a $\beta$-Sn-equivalent structure upon pressurization: the transition geometry would result in the nearing of like-charged, mutually repulsive atoms \cite{Ackland2001a}.
Most III-V semiconducting zincblende-formers transition to wurtzite, a hexagonal diamond derivative under pressure. While $\beta$-Sn is topologically identical to diamond, wurtzite is to cubic diamond as cubic-close sphere packing is to its hexagonally-close packed relative: the local coordination is identical, but its network is topologically distinct, which would require a breaking of bonds and rearrangement of the constituent atoms in the transition.


We have predicted a new phase transition achievable in colloidal systems, induced by a slight change in pressure or in particle shape.
This structural transition provides the exciting prospect of making photonic materials switchable between multiple photonic band gap states, either between two different gaps, or between ``on''/``off'' states.
Given the placement of the two possible gaps, complete photonic band gaps would be achievable in colloidal crystals in visible or infrared wavelengths, and could potentially toggle between two colors or separate regimes in the electromagnetic spectrum.
For example, using dielectric spheres with $\epsilon=11.56$ (silicon) that inscribe truncated tetrahedra with $\alpha=0.5$ (see Fig.~{\ref{fig:photonic}) and a lattice constant of $a=184~\text{nm}$, a three-state material can be assembled with a violet ($400~\text{nm}$) band gap at low densities, no band gap at intermediate densities, and a red ($670~\text{nm}$) band gap at high densities.
If a larger lattice constant of $a=8.2~\mu\text{m}$ is used, which is achievable for colloidal systems, the lower band gap occurs at $30~\mu\text{m}$, or about $10~\text{THz}$, suitable for sub-millimeter photonic applications.
If $a$ is a tunable design parameter, this system can exhibit bands gaps at target wavelengths whose ratios are approximately $1.67$.
The colloidal crystal structures examined should be obtainable through other means, such as DNA-programmable assembly.  The lengths and bonding strength of DNA needed to achieve each structure may be gleaned by examination of the PMFTs \cite{Liu2016,Mirkin}.
In practice, it would be important to remove thermal disorder from the colloidal crystals before and after the transition, e.g. by quenching, as disorder is known to destroy photonic band gaps.

\section{Acknowledgements}
We thank P.\ F.\ Damasceno for helpful conversations toward calculating the photonic properties of our structures.
We thank C.\ X.\ Du for valuable discussions.
We thank E.\ S.\ Harper for help with visualizing the PMFTs.
We thank S.\ C.\ Barterian and B.D.~Dice for giving useful feedback during the writing process.
We also thank K.\ A.\ Mengle, N.\ Sanders, and E.\ Kioupakis for guidance with calculations.

R.\ K.\ C.\  acknowledges support by the National Science Foundation, Division of Materials Research Award \# DMR 1120923 and by the University of Michigan Rackham Merit Fellowship program. This material is based upon work supported by a Simons Investigator Award from the Simons Foundation to S.\ C.\ G.\, and a National Science Foundation Graduate Research Fellowship Grant No. DGE 1256260 to J.\ A.\. This work used the Extreme Science and Engineering Discovery Environment (XSEDE), which is supported by National Science Foundation grant number ACI-1053575; XSEDE award \# DMR 140129. Computational resources and services were also provided by Advanced Research Computing at the University of Michigan, Ann Arbor.


\begin{thebibliography}{29}%
\makeatletter
\providecommand \@ifxundefined [1]{%
 \@ifx{#1\undefined}
}%
\providecommand \@ifnum [1]{%
 \ifnum #1\expandafter \@firstoftwo
 \else \expandafter \@secondoftwo
 \fi
}%
\providecommand \@ifx [1]{%
 \ifx #1\expandafter \@firstoftwo
 \else \expandafter \@secondoftwo
 \fi
}%
\providecommand \natexlab [1]{#1}%
\providecommand \enquote  [1]{``#1''}%
\providecommand \bibnamefont  [1]{#1}%
\providecommand \bibfnamefont [1]{#1}%
\providecommand \citenamefont [1]{#1}%
\providecommand \href@noop [0]{\@secondoftwo}%
\providecommand \href [0]{\begingroup \@sanitize@url \@href}%
\providecommand \@href[1]{\@@startlink{#1}\@@href}%
\providecommand \@@href[1]{\endgroup#1\@@endlink}%
\providecommand \@sanitize@url [0]{\catcode `\\12\catcode `\$12\catcode
  `\&12\catcode `\#12\catcode `\^12\catcode `\_12\catcode `\%12\relax}%
\providecommand \@@startlink[1]{}%
\providecommand \@@endlink[0]{}%
\providecommand \url  [0]{\begingroup\@sanitize@url \@url }%
\providecommand \@url [1]{\endgroup\@href {#1}{\urlprefix }}%
\providecommand \urlprefix  [0]{URL }%
\providecommand \Eprint [0]{\href }%
\providecommand \doibase [0]{http://dx.doi.org/}%
\providecommand \selectlanguage [0]{\@gobble}%
\providecommand \bibinfo  [0]{\@secondoftwo}%
\providecommand \bibfield  [0]{\@secondoftwo}%
\providecommand \translation [1]{[#1]}%
\providecommand \BibitemOpen [0]{}%
\providecommand \bibitemStop [0]{}%
\providecommand \bibitemNoStop [0]{.\EOS\space}%
\providecommand \EOS [0]{\spacefactor3000\relax}%
\providecommand \BibitemShut  [1]{\csname bibitem#1\endcsname}%
\let\auto@bib@innerbib\@empty
\bibitem [{\citenamefont {Bragg}\ and\ \citenamefont
  {Bragg}(1913)}]{Bragg1913}%
  \BibitemOpen
  \bibfield  {author} {\bibinfo {author} {\bibfnamefont {W.~H.}\ \bibnamefont
  {Bragg}}\ and\ \bibinfo {author} {\bibfnamefont {W.~L.}\ \bibnamefont
  {Bragg}},\ }\href {\doibase 10.1038/091557a0} {\bibfield  {journal} {\bibinfo
   {journal} {Nature}\ }\textbf {\bibinfo {volume} {91}},\ \bibinfo {pages}
  {557} (\bibinfo {year} {1913})}\BibitemShut {NoStop}%
\bibitem [{\citenamefont {Ho}\ \emph {et~al.}(1990)\citenamefont {Ho},
  \citenamefont {Chan}, \citenamefont {Soukoulis},\ and\ \citenamefont
  {Soukoulis}}]{Ho1990}%
  \BibitemOpen
  \bibfield  {author} {\bibinfo {author} {\bibfnamefont {K.~M.}\ \bibnamefont
  {Ho}}, \bibinfo {author} {\bibfnamefont {C.~T.}\ \bibnamefont {Chan}},
  \bibinfo {author} {\bibnamefont {Soukoulis}}, \ and\ \bibinfo {author}
  {\bibfnamefont {C.~M.}\ \bibnamefont {Soukoulis}},\ }\href {\doibase
  10.1103/PhysRevLett.65.3152} {\bibfield  {journal} {\bibinfo  {journal}
  {Phys. Rev. Lett.}\ }\textbf {\bibinfo {volume} {65}},\ \bibinfo {pages}
  {3152--3155} (\bibinfo {year} {1990})}\BibitemShut {NoStop}%
\bibitem [{\citenamefont {Ackland}(2001{\natexlab{a}})}]{Ackland2001}%
  \BibitemOpen
  \bibfield  {author} {\bibinfo {author} {\bibfnamefont {G.~J.}\ \bibnamefont
  {Ackland}},\ }\href {\doibase 10.1088/0034-4885/64/4/202} {\bibfield
  {journal} {\bibinfo  {journal} {Reports on Progress in Physics}\ }\textbf
  {\bibinfo {volume} {64}},\ \bibinfo {pages} {483--516} (\bibinfo {year}
  {2001}{\natexlab{a}})}\BibitemShut {NoStop}%
\bibitem [{\citenamefont {Clark}\ \emph {et~al.}(1994)\citenamefont {Clark},
  \citenamefont {Ackland},\ and\ \citenamefont {Crain}}]{Clark1994}%
  \BibitemOpen
  \bibfield  {author} {\bibinfo {author} {\bibfnamefont {S.~J.}\ \bibnamefont
  {Clark}}, \bibinfo {author} {\bibfnamefont {G.~J.}\ \bibnamefont {Ackland}},
  \ and\ \bibinfo {author} {\bibfnamefont {J.}~\bibnamefont {Crain}},\ }\href
  {\doibase 10.1103/PhysRevB.49.5341} {\bibfield  {journal} {\bibinfo
  {journal} {Physical Review B}\ }\textbf {\bibinfo {volume} {49}},\ \bibinfo
  {pages} {5341--5352} (\bibinfo {year} {1994})}\BibitemShut {NoStop}%
\bibitem [{\citenamefont {McMillan}\ \emph {et~al.}(2005)\citenamefont
  {McMillan}, \citenamefont {Wilson}, \citenamefont {Daisenberger},\ and\
  \citenamefont {Machon}}]{Mcmillan2005}%
  \BibitemOpen
  \bibfield  {author} {\bibinfo {author} {\bibfnamefont {P.~F.}\ \bibnamefont
  {McMillan}}, \bibinfo {author} {\bibfnamefont {M.}~\bibnamefont {Wilson}},
  \bibinfo {author} {\bibfnamefont {D.}~\bibnamefont {Daisenberger}}, \ and\
  \bibinfo {author} {\bibfnamefont {D.}~\bibnamefont {Machon}},\ }\href
  {\doibase 10.1038/nmat1458} {\bibfield  {journal} {\bibinfo  {journal}
  {Nature Materials}\ }\textbf {\bibinfo {volume} {4}},\ \bibinfo {pages} {680--684}
  (\bibinfo {year} {2005})}\BibitemShut {NoStop}%
\bibitem [{\citenamefont {Jamieson}(1963)}]{Jamieson1963}%
  \BibitemOpen
  \bibfield  {author} {\bibinfo {author} {\bibfnamefont {J.~C.}\ \bibnamefont
  {Jamieson}},\ }\href {\doibase 10.1126/science.139.3556.762} {\bibfield
  {journal} {\bibinfo  {journal} {Science}\ }\textbf {\bibinfo {volume}
  {139}},\ \bibinfo {pages} {762--764} (\bibinfo {year} {1963})}\BibitemShut
  {NoStop}%
\bibitem [{\citenamefont {Bundy}\ \emph {et~al.}(1996)\citenamefont {Bundy},
  \citenamefont {Bassett}, \citenamefont {Weathers}, \citenamefont {Hemley},
  \citenamefont {Mao},\ and\ \citenamefont {Goncharov}}]{Bundy1996}%
  \BibitemOpen
  \bibfield  {author} {\bibinfo {author} {\bibfnamefont {F.~P.}\ \bibnamefont
  {Bundy}}, \bibinfo {author} {\bibfnamefont {W.~A.}\ \bibnamefont {Bassett}},
  \bibinfo {author} {\bibfnamefont {M.~S.}\ \bibnamefont {Weathers}}, \bibinfo
  {author} {\bibfnamefont {R.~J.}\ \bibnamefont {Hemley}}, \bibinfo {author}
  {\bibfnamefont {H.~K.}\ \bibnamefont {Mao}}, \ and\ \bibinfo {author}
  {\bibfnamefont {A.~F.}\ \bibnamefont {Goncharov}},\ }\href {\doibase
  10.1016/0008-6223(96)00170-4} {\bibfield  {journal} {\bibinfo  {journal}
  {Carbon}\ }\textbf {\bibinfo {volume} {34}},\ \bibinfo {pages} {141--153}
  (\bibinfo {year} {1996})}\BibitemShut {NoStop}%
\bibitem [{\citenamefont {Katzke}\ \emph {et~al.}(2006)\citenamefont {Katzke},
  \citenamefont {Bismayer},\ and\ \citenamefont {Tol{\'{e}}dano}}]{Katzke2006}%
  \BibitemOpen
  \bibfield  {author} {\bibinfo {author} {\bibfnamefont {H.}~\bibnamefont
  {Katzke}}, \bibinfo {author} {\bibfnamefont {U.}~\bibnamefont {Bismayer}}, \
  and\ \bibinfo {author} {\bibfnamefont {P.}~\bibnamefont {Tol{\'{e}}dano}},\
  }\href {\doibase 10.1103/PhysRevB.73.134105} {\bibfield  {journal} {\bibinfo
  {journal} {Physical Review B}\ }\textbf {\bibinfo {volume} {73}},\ \bibinfo
  {pages} {134105} (\bibinfo {year} {2006})}\BibitemShut {NoStop}%
\bibitem [{\citenamefont {Ackland}(2001{\natexlab{b}})}]{Ackland2001a}%
  \BibitemOpen
  \bibfield  {author} {\bibinfo {author} {\bibfnamefont {G.}~\bibnamefont
  {Ackland}},\ }\href {\doibase
  10.1002/1521-3951(200101)223:2<361::AID-PSSB361>3.0.CO;2-1} {\bibfield
  {journal} {\bibinfo  {journal} {physica status solidi (b)}\ }\textbf
  {\bibinfo {volume} {223}},\ \bibinfo {pages} {361--368} (\bibinfo {year}
  {2001}{\natexlab{b}})}\BibitemShut {NoStop}%
\bibitem [{\citenamefont {Goncharov}(1987)}]{Goncharov1987}%
  \BibitemOpen
  \bibfield  {author} {\bibinfo {author} {\bibfnamefont {A.~F.}\ \bibnamefont
  {Goncharov}},\ }\href {\doibase 10.1070/PU1987v030n06ABEH002854} {\bibfield
  {journal} {\bibinfo  {journal} {Soviet Physics Uspekhi}\ }\textbf {\bibinfo
  {volume} {30}},\ \bibinfo {pages} {525--534} (\bibinfo {year} {1987})}\BibitemShut
  {NoStop}%
\bibitem [{\citenamefont {Zhang}\ \emph {et~al.}(2005)\citenamefont {Zhang},
  \citenamefont {Keys}, \citenamefont {Chen},\ and\ \citenamefont
  {Glotzer}}]{Zhang2005}%
  \BibitemOpen
  \bibfield  {author} {\bibinfo {author} {\bibfnamefont {Z.}~\bibnamefont
  {Zhang}}, \bibinfo {author} {\bibfnamefont {A.~S.}\ \bibnamefont {Keys}},
  \bibinfo {author} {\bibfnamefont {T.}~\bibnamefont {Chen}}, \ and\ \bibinfo
  {author} {\bibfnamefont {S.~C.}\ \bibnamefont {Glotzer}},\ }\href {\doibase
  10.1021/la0513611} {\bibfield  {journal} {\bibinfo  {journal} {Langmuir}\ }\textbf {\bibinfo {volume} {21}},\
  \bibinfo {pages} {11547--11551} (\bibinfo {year} {2005})}\BibitemShut {NoStop}%
\bibitem [{\citenamefont {Ducrot}\ \emph {et~al.}(2017)\citenamefont {Ducrot},
  \citenamefont {He}, \citenamefont {Yi},\ and\ \citenamefont
  {Pine}}]{Ducrot2017}%
  \BibitemOpen
  \bibfield  {author} {\bibinfo {author} {\bibfnamefont {{\'{E}}.}~\bibnamefont
  {Ducrot}}, \bibinfo {author} {\bibfnamefont {M.}~\bibnamefont {He}}, \bibinfo
  {author} {\bibfnamefont {G.-R.}\ \bibnamefont {Yi}}, \ and\ \bibinfo {author}
  {\bibfnamefont {D.~J.}\ \bibnamefont {Pine}},\ }\href {\doibase
  10.1038/nmat4869} {\bibfield  {journal} {\bibinfo  {journal} {Nature
  Materials}\ }\textbf {\bibinfo {volume} {16}},\ \bibinfo {pages} {652--657}
  (\bibinfo {year} {2017})}\BibitemShut {NoStop}%
\bibitem [{\citenamefont {Damasceno}\ \emph {et~al.}(2012)\citenamefont
  {Damasceno}, \citenamefont {Engel},\ and\ \citenamefont
  {Glotzer}}]{Damasceno2012d}%
  \BibitemOpen
  \bibfield  {author} {\bibinfo {author} {\bibfnamefont {P.~F.}\ \bibnamefont
  {Damasceno}}, \bibinfo {author} {\bibfnamefont {M.}~\bibnamefont {Engel}}, \
  and\ \bibinfo {author} {\bibfnamefont {S.~C.}\ \bibnamefont {Glotzer}},\
  }\href {\doibase 10.1021/nn204012y} {\bibfield  {journal} {\bibinfo
  {journal} {ACS Nano}\ }\textbf {\bibinfo {volume} {6}},\ \bibinfo {pages}
  {609--614} (\bibinfo {year} {2012})}\BibitemShut {NoStop}%
\bibitem [{\citenamefont {Jain}\ \emph {et~al.}(2014)\citenamefont {Jain},
  \citenamefont {Errington},\ and\ \citenamefont {Truskett}}]{Jain}%
  \BibitemOpen
  \bibfield  {author} {\bibinfo {author} {\bibfnamefont {A.}~\bibnamefont
  {Jain}}, \bibinfo {author} {\bibfnamefont {J.~R.}\ \bibnamefont {Errington}},
  \ and\ \bibinfo {author} {\bibfnamefont {T.~M.}\ \bibnamefont {Truskett}},\
  }\href {\doibase 10.1103/PhysRevX.4.031049} {\bibfield  {journal} {\bibinfo
  {journal} {Physical Review X}\ }\textbf {\bibinfo {volume} {4}}, \bibinfo {pages} {049902} (\bibinfo
  {year} {2014})}\BibitemShut {NoStop}%
\bibitem [{\citenamefont {Wang}\ \emph {et~al.}(2017)\citenamefont {Wang},
   \citenamefont {Jenkins},  \citenamefont {McGinley}, \citenamefont {Sinno},\ and\ \citenamefont {Crocker}}]{Wang}%
  \BibitemOpen
  \bibfield  {author} {\bibinfo {author} {\bibfnamefont {Y.}~\bibnamefont
  {Wang}}, \bibinfo {author} {\bibfnamefont {I.~C.}\ \bibnamefont {Jenkins}},
   \bibinfo {author} {\bibfnamefont {J.~T.}\ \bibnamefont {McGinley}},
    \bibinfo {author} {\bibfnamefont {T.}\ \bibnamefont {Sinno}}, and\ \bibinfo {author} {\bibfnamefont {J.~C.}\ \bibnamefont {Crocker}},\
  }\href {\doibase 10.1038/ncomms14173} {\bibfield  {journal} {\bibinfo
  {journal} {Nature Communications}\ }\textbf {\bibinfo {volume} {8}}, \bibinfo {pages} {14173} (\bibinfo
  {year} {2017})}\BibitemShut {NoStop}%
 \bibitem [{\citenamefont {Romano2010}\ \emph {et~al.}(2010)\citenamefont {Romano},
   \citenamefont {Sanz},  and\ \citenamefont {Sciotino}}]{Romano2010}%
  \BibitemOpen
  \bibfield  {author} {\bibinfo {author} {\bibfnamefont {F.}~\bibnamefont
  {Romano}}, \bibinfo {author} {\bibfnamefont {E.}\ \bibnamefont {Sanz}},
  and\ \bibinfo {author} {\bibfnamefont {F.}\ \bibnamefont {Sciortino}},\
  }\href {\doibase 10.1063/1.3393777} {\bibfield  {journal} {\bibinfo
  {journal} {Journal of Chemical Physics}\ }\textbf {\bibinfo {volume} {132}}, \bibinfo {pages} {184501} (\bibinfo
  {year} {2010})}\BibitemShut {NoStop}%
\bibitem{Chen2014}
\bibinfo{author}{E.~R. Chen}, \bibinfo{author}{D. Klotsa},
  \bibinfo{author}{M. Engel}, \bibinfo{author}{P.~F. Damasceno} \&
  \bibinfo{author}{S.~C. Glotzer}
\newblock \emph{\bibinfo{journal}{Phys. Rev. X}} \textbf{\bibinfo{volume}{4}},
  \bibinfo{pages}{11024} (\bibinfo{year}{2014}).
\bibitem [{\citenamefont {van Anders}\ \emph {et~al.}(2015)\citenamefont {van
  Anders}, \citenamefont {Klotsa}, \citenamefont {Karas}, \citenamefont
  {Dodd},\ and\ \citenamefont {Glotzer}}]{vanAnders2015}%
  \BibitemOpen
  \bibfield  {author} {\bibinfo {author} {\bibfnamefont {G.}~\bibnamefont {van
  Anders}}, \bibinfo {author} {\bibfnamefont {D.}~\bibnamefont {Klotsa}},
  \bibinfo {author} {\bibfnamefont {A.~S.}\ \bibnamefont {Karas}}, \bibinfo
  {author} {\bibfnamefont {P.~M.}\ \bibnamefont {Dodd}}, \ and\ \bibinfo
  {author} {\bibfnamefont {S.~C.}\ \bibnamefont {Glotzer}},\ }\href {\doibase
  10.1021/acsnano.5b04181} {\bibfield  {journal} {\bibinfo  {journal} {ACS
  Nano}\ }\textbf {\bibinfo {volume} {9}},\ \bibinfo {pages} {9542--9553} (\bibinfo
  {year} {2015})}\BibitemShut {NoStop}%
\bibitem [{\citenamefont {Cersonsky}\ \emph {et~al.}(2017)\citenamefont
  {Cersonsky}, \citenamefont {van Anders}, \citenamefont {Dodd},\ and\
  \citenamefont {Glotzer}}]{Cersonsky2017}%
  \BibitemOpen
  \bibfield  {author} {\bibinfo {author} {\bibfnamefont {R.}~\bibnamefont
  {Cersonsky}}, \bibinfo {author} {\bibfnamefont {G.}~\bibnamefont {van
  Anders}}, \bibinfo {author} {\bibfnamefont {P.~M.}\ \bibnamefont {Dodd}}, \
  and\ \bibinfo {author} {\bibfnamefont {S.~C.}\ \bibnamefont {Glotzer}},\
  }\href {\doibase
  10.1021/pnas.1720139115} {\bibfield  {journal} {\bibinfo  {journal}{Proceedings of the National Academy of Science}}} \BibitemShut {NoStop}%
\bibitem [{\citenamefont {Ehrenfest}(1933)}]{Ehrenfest}%
  \BibitemOpen
  \bibfield  {author} {\bibinfo {author} {\bibfnamefont {P.}~\bibnamefont
  {Ehrenfest}},\ }\href@noop {} {\bibfield  {journal} {\bibinfo  {journal}
  {Proceedings of the Koninklijke Akademie Van
  Wetenschappen Te Amsterdam}\ }\textbf {\bibinfo {volume} {36}},\ \bibinfo
  {pages} {153--157} (\bibinfo {year} {1933})}\BibitemShut {NoStop}%
\bibitem [{\citenamefont {van Anders}\ \emph {et~al.}(2014)\citenamefont {van
  Anders}, \citenamefont {Klotsa}, \citenamefont {Ahmed}, \citenamefont
  {Engel},\ and\ \citenamefont {Glotzer}}]{VanAnders2014d}%
  \BibitemOpen
  \bibfield  {author} {\bibinfo {author} {\bibfnamefont {G.}~\bibnamefont {van
  Anders}}, \bibinfo {author} {\bibfnamefont {D.}~\bibnamefont {Klotsa}},
  \bibinfo {author} {\bibfnamefont {N.~K.}\ \bibnamefont {Ahmed}}, \bibinfo
  {author} {\bibfnamefont {M.}~\bibnamefont {Engel}}, \ and\ \bibinfo {author}
  {\bibfnamefont {S.~C.}\ \bibnamefont {Glotzer}},\ }\href {\doibase
  10.1073/pnas.1418159111} {\bibfield  {journal} {\bibinfo  {journal}
  {Proceedings of the National Academy of Sciences}\ }\textbf {\bibinfo
  {volume} {111}},\ \bibinfo {pages} {E4812--E4821} (\bibinfo {year}
  {2014})}\BibitemShut {NoStop}%
  \bibitem [{\citenamefont {Johnson}\ and\ \citenamefont
  {Joannopoulos}(2001)}]{Johnson2001}%
  \BibitemOpen
  \bibfield  {author} {\bibinfo {author} {\bibfnamefont {S.~G.}\ \bibnamefont
  {Johnson}}\ and\ \bibinfo {author} {\bibfnamefont {J.}~\bibnamefont
  {Joannopoulos}},\ }\href
  {http://www.opticsexpress.org/abstract.cfm?URI=OPEX-8-3-173} {\bibfield
  {journal} {\bibinfo  {journal} {Opt. Express}\ }\textbf {\bibinfo {volume}
  {8}},\ \bibinfo {pages} {173--190} (\bibinfo {year} {2001})}\BibitemShut {NoStop}%
\bibitem [{\citenamefont {Li}\ and\ \citenamefont {Liu}(2017)}]{Li2017}%
  \BibitemOpen
  \bibfield  {author} {\bibinfo {author} {\bibfnamefont {Y.}~\bibnamefont
  {Li}}\ and\ \bibinfo {author} {\bibfnamefont {B.}~\bibnamefont {Liu}},\
  }\href {\doibase 10.1021/acsmacrolett.7b00629} {\bibfield  {journal}
  {\bibinfo  {journal} {ACS Macro Letters}\ }\textbf {\bibinfo {volume} {6}},\
  \bibinfo {pages} {1315--1319} (\bibinfo {year} {2017})}\BibitemShut {NoStop}%
\bibitem [{\citenamefont {Liu}\ \emph {et~al.}(2016)\citenamefont {Liu},
  \citenamefont {Tagawa}, \citenamefont {Xin}, \citenamefont {Wang},
  \citenamefont {Emamy}, \citenamefont {Li}, \citenamefont {Yager},
  \citenamefont {Starr}, \citenamefont {Tkachenko},\ and\ \citenamefont
  {Gang}}]{Liu2016}%
  \BibitemOpen
  \bibfield  {author} {\bibinfo {author} {\bibfnamefont {W.}~\bibnamefont
  {Liu}}, \bibinfo {author} {\bibfnamefont {M.}~\bibnamefont {Tagawa}},
  \bibinfo {author} {\bibfnamefont {H.~L.}\ \bibnamefont {Xin}}, \bibinfo
  {author} {\bibfnamefont {T.}~\bibnamefont {Wang}}, \bibinfo {author}
  {\bibfnamefont {H.}~\bibnamefont {Emamy}}, \bibinfo {author} {\bibfnamefont
  {H.}~\bibnamefont {Li}}, \bibinfo {author} {\bibfnamefont {K.~G.}\
  \bibnamefont {Yager}}, \bibinfo {author} {\bibfnamefont {F.~W.}\ \bibnamefont
  {Starr}}, \bibinfo {author} {\bibfnamefont {A.~V.}\ \bibnamefont
  {Tkachenko}}, \ and\ \bibinfo {author} {\bibfnamefont {O.}~\bibnamefont
  {Gang}},\ }\href {\doibase 10.1126/science.aad2080} {\bibfield  {journal}
  {\bibinfo  {journal} {Science}\ }\textbf {\bibinfo {volume}
  {351}},\ \bibinfo {pages} {582--586} (\bibinfo {year} {2016})}\BibitemShut
  {NoStop}%
 \bibitem [{\citenamefont {Lin}\ \emph {et~al.}(2017)\citenamefont
 Lin H1,2, Lee S3, Sun L2,4, Spellings M3, Engel M5, Glotzer SC6,7,8, Mirkin CA9,2,4.
  {Lee}, \citenamefont {Sun},\citenamefont {Spellings}, \citenamefont {Engel}, \citenamefont {Glotzer},
  and\ \citenamefont {Mirkin}}]{Mirkin}%
  \BibitemOpen
  \bibfield  {author} {\bibinfo {author} {\bibfnamefont {H.}\ \bibnamefont
  {Lin}}, \bibinfo {author} {\bibfnamefont {S.}\ \bibnamefont
  {Sun}}, 
  \bibinfo {author} {\bibfnamefont {M.}\ \bibnamefont
  {Spellings}},
  \bibinfo {author} {\bibfnamefont {M.}\ \bibnamefont
  {Engel}},
  \bibinfo {author} {\bibfnamefont {S.~C.}\ \bibnamefont
  {Glotzer}},
  \ and\ \bibinfo {author} {\bibfnamefont {C.~A.}~\bibnamefont
  {Mirkin}},\ }\href {\doibase 10.1016/j.jcp.2008.01.047} {\bibfield
  {journal} {\bibinfo  {journal} {Science}\ }\textbf
  {\bibinfo {volume} {355}},\ \bibinfo {pages} {931-935} (\bibinfo {year}
  {2017})}\BibitemShut {NoStop}%
\bibitem [{\citenamefont {Anderson}\ \emph {et~al.}(2008)\citenamefont
  {Anderson}, \citenamefont {Lorenz},\ and\ \citenamefont
  {Travesset}}]{anderson2008}%
  \BibitemOpen
  \bibfield  {author} {\bibinfo {author} {\bibfnamefont {J.~A.}\ \bibnamefont
  {Anderson}}, \bibinfo {author} {\bibfnamefont {C.~D.}\ \bibnamefont
  {Lorenz}}, \ and\ \bibinfo {author} {\bibfnamefont {A.}~\bibnamefont
  {Travesset}},\ }\href {\doibase 10.1016/j.jcp.2008.01.047} {\bibfield
  {journal} {\bibinfo  {journal} {Journal of Computational Physics}\ }\textbf
  {\bibinfo {volume} {227}},\ \bibinfo {pages} {5342--5359} (\bibinfo {year}
  {2008})}\BibitemShut {NoStop}%
\bibitem [{\citenamefont {Anderson}\ \emph {et~al.}(2016)\citenamefont
  {Anderson}, \citenamefont {{Eric Irrgang}},\ and\ \citenamefont
  {Glotzer}}]{anderson2016}%
  \BibitemOpen
  \bibfield  {author} {\bibinfo {author} {\bibfnamefont {J.~A.}\ \bibnamefont
  {Anderson}}, \bibinfo {author} {\bibfnamefont {M.}~\bibnamefont {{Eric
  Irrgang}}}, \ and\ \bibinfo {author} {\bibfnamefont {S.~C.}\ \bibnamefont
  {Glotzer}},\ }\href {\doibase 10.1016/j.cpc.2016.02.024} {\bibfield
  {journal} {\bibinfo  {journal} {Computer Physics Communications}\ }\textbf
  {\bibinfo {volume} {204}},\ \bibinfo {pages} {21--30} (\bibinfo {year}
  {2016})}\BibitemShut {NoStop}%
\bibitem [{\citenamefont {Adorf}\ \emph {et~al.}(2016)\citenamefont {Adorf},
  \citenamefont {Dodd},\citenamefont {Ramasubramani}, \ and\ \citenamefont {Glotzer}}]{Adorf2016}%
  \BibitemOpen
  \bibfield  {author} {\bibinfo {author} {\bibfnamefont {C.~S.}\ \bibnamefont
  {Adorf}}, \bibinfo {author} {\bibfnamefont {P.~M.}\ \bibnamefont {Dodd}}, \bibinfo {author} {\bibfnamefont {V.}\ \bibnamefont {Ramasubramani}}, \
  and\ \bibinfo {author} {\bibfnamefont {S.~C.}\ \bibnamefont {Glotzer}},\
  }{\bibfield  {journal} \href {\doibase 10.1016/j.commatsci.2018.01.035} {\bibinfo
  {journal} {Comp. Mater. Sci.}\ }\textbf {\bibinfo {volume} {146}} (\bibinfo {year}
  {2018})}
  \BibitemShut {NoStop}%
\bibitem{seekpath}
\bibinfo{author}{Hinuma, Y.}, \bibinfo{author}{Pizzi, G.},
  \bibinfo{author}{Kumagai, Y.}, \bibinfo{author}{Oba, F.} \&
  \bibinfo{author}{Tanaka, I.}
\newblock \emph{\bibinfo{journal}{Computational Materials Science}}
  \textbf{\bibinfo{volume}{128}}, \bibinfo{pages}{140 -- 184}
  (\bibinfo{year}{2017}).

\end{thebibliography}

\appendix

\section{Model and Methods}
We conduct our study in the alchemical thermodynamic ensemble, implemented through the Digital Alchemy framework \cite{vanAnders2015}.
This ensemble is an extension of the canonical ensemble that treats particle shape, parametrized by a set of values $\alpha_i$, as a thermodynamic state variable.
Each $\alpha_i$ parameter has an associated conjugate \textit{alchemical potential} $\mu_i$.
The $NVT\mu$ ensemble, in which each $\alpha_i$ is allowed to fluctuate, is defined with the partition function

\begin{equation}
Z = \sum_\sigma e^{-\beta\left(H-\sum_i \mu_i N \alpha_i\right)} .
\label{eq:partition}
\end{equation}

Here, $\sum_\sigma$ denotes a sum over all microstates $\sigma$ and shape parameter values $\alpha_i$.
The variables $\alpha_i$ and $\mu_i$ are related to derivatives of $F$, the thermodynamic potential for the ensemble:

\begin{equation}
\mu_i = \frac{1} {N} \left(\frac{\partial F} {\partial \alpha_i} \right)_{N,\phi,T,\alpha_{j\neq i} }
\end{equation}
and
\begin{equation}
\alpha_i = \frac{1} {N} \left(\frac{\partial F} {\partial \mu_i} \right)_{N,\phi,T,\mu_{j\neq i} },
\end{equation}

\noindent where

\begin{equation}
F = -k_B T \log\left(Z\right).
\end{equation}

For all simulations, we used the hard-particle Monte Carlo sampling method of the HOOMD-blue simulation toolkit \cite{anderson2008,anderson2016}.
We conducted alchemical hard particle Monte Carlo (Alch-HPMC) simulations within the $NVT\alpha$, $NPT\alpha$, and $NVT\mu$ ensembles. Further details on Alch-HPMC can be found in Refs.~\cite{vanAnders2015, Cersonsky2017}.
All simulations were run so that each particle's shape is specified by one $\alpha$ and every particle in the system changes identically and simultaneously as $\alpha$ changes.

\subsection{Simulations in $NVT\alpha$}
In the $NVT\alpha$ ensemble, we simulated systems of 1728 particles with unit volume, with state points from $\alpha \in [0.3, 0.5]$ with $\Delta\alpha = 0.01$ initialized in the diamond structure at $\phi = 0.55$, the known assembly structure for these shapes \cite{Damasceno2012d}.
These simulations were then incrementally compressed to a target density of $\phi = 0.55$--$0.95$ with $\Delta\phi = 0.05$, allowing for the box aspect ratio to change as a separate MC move while maintaining constant box volume.
For regions of interest, additional state points were run at $\Delta\phi = 0.025$.
The box aspect ratio search radius was set to $0.01$ per sweep.
Simulations were then run for $1.1\times10^7$--$2.4\times10^7$ MC sweeps, until equilibrated.
A minimum of four replicas were run for each state point.

Similar simulations were run by initializing at the densest packing of the shape family in both diamond and its tetragonal derivative.
The box was expanded isotropically to a target density, then simulated using MC with box moves for $1.1\times10^7$--$2.4\times10^7$ sweeps.
These simulations were run with the same parameters as systems initialized in the diamond assembly phase.

\subsubsection{Analysis}

Each frame of a simulation was separated into crystal grains using in-house environment-matching algorithms (E.\ G.\ Teich, \textit{private communication}) analyzing the bond angles of each particle with its nearest neighbors and then separating the particles into groups based upon these arrangements.
For some simulations, compression and the reduction in symmetry caused the formation of multiple grains within the assembly, and these state points were removed from the results to avoid distorting the influence of these simulation data.
We used in-house software \textit{injavis} to determine the 8-particle unit cells of all simulation runs.
We used the signac framework for all data management \cite{Adorf2016}.

\subsection{Simulations in $NPT\alpha$}
We simulated systems of 512 particles with unit volume and constant shape in an $NPT\alpha$ ensemble to generate an equation of state for the given transformation at 35 unitless pressures ($=\frac{PV} {k_BT} $ with unit volume) between $\beta P=5$--$100$ and $\alpha = 0.4$--$0.5$, with $\Delta\alpha = 0.01$.
Additional state points were run for regions of interest and for $\alpha=0.5$.
Systems were initialized in the tetragonal diamond derivative ($c/a = \sqrt{0.4} $) at maximum density or in their assembled diamond phase at the assembly density, and held at target pressure to equilibrate over $1\times 10^7$--$4\times 10^7$ MC sweeps. Simulations for shapes of interest ($\alpha=0.5, P=9.8$--$10.0$) were run with longer simulations and additional state points (up to 10). A minimum of three independent simulations were run for all other state points.

\section{Potential of Mean Force and Torque (PMFT)}

For the data shown, we averaged the PMFT values for 150 frames from earlier $NVT\mu$ simulations at a range of $\phi$ values.
The PMFT results were visualized using the Python package \textit{mayavi}.

\subsection{Simulations in $NVT\mu$}

Alch-HPMC simulations were run in the $NVT\mu$ ensemble to find the thermodynamically preferred structure and particle shape.
Systems of 512 particles were initialized with $\alpha = \frac{1} {3} $ and $\phi = 0.6$ and allowed to equilibrate in shape space.
The particles were then compressed isotropically to a target $\phi \in [0.70, 0.95]$ with $\Delta\phi = 0.05$, and allowed to equilibrate using MC box moves and subsequently within shape space.
This step was repeated until both structure and shape were at equilibrium.
Systems were then decompressed to $\phi = 0.60$ and allowed once again to equilibrate to check for reversibility.
A diagram of the simulation protocol can be found in the Supplementary Information.

\section{Calculation of Photonic Band Structure}

We computed the photonic band structure for different forms of tetragonal diamond using MIT Photonic Bands (MPB) \cite{Johnson2001}. This package computes the eigenmodes of Maxwell's equations through conjugate-gradient minimization of the block Rayleigh quotient in a plane-wave basis \cite{Johnson2001}.

We computed the size and location of all photonic band gaps for all structures computed in $NVT\alpha$ simulations in Fig.~{\ref{fig:panel2}\textbf{(b)}, including both $c/a$ ratio and lattice vector magnitude (to reflect density). For radius we took the insphere of the respective polyhedra in $\Delta_{323}$ and for the dielectric constant we chose 11.56, that of silicon. We computed the irreducible Brillouin zone using the package \textit{SeeK-path} \cite{seekpath}, an open-source k-space path finder.

Resulting photonic band gaps were normalized by the mid-gap frequency, consistent with convention.

\end{document}